\documentclass[reprint,article,superscriptaddress,msmath,amssymb,aps]{revtex4-1}
\usepackage{graphicx}
\usepackage{dcolumn}
\usepackage{bm}
\usepackage{color}
\usepackage[]{natbib}
\usepackage{amsmath,amssymb,amsfonts,mathtools}
\usepackage{float}
\usepackage{pdfcomment}
\usepackage{verbatim}
\usepackage[utf8]{inputenc}
\usepackage{booktabs,chemformula}

\begin{document}
\preprint{APS/123-QED}
\title{Structural, Elastic and Electronic Properties of SmFeO\textsubscript{3} using Density Functional Theory}

\author{Shahran Ahmed}
\affiliation{Department of Electrical and Electronic Engineering, University of Dhaka, Dhaka-1000, Bangladesh}

\author{Sadiq Shahriyar Nishat}
\affiliation{Department of Physics, University of Dhaka, Dhaka-1000, Bangladesh}

\author{Alamgir Kabir}
\affiliation{Department of Physics, University of Dhaka, Dhaka-1000, Bangladesh}

\author{A. K. M. Sarwar Hossain Faysal}
\affiliation{Department of Electrical and Electronic Engineering, University of Dhaka, Dhaka-1000, Bangladesh}

\author{Tarique Hasan}
\affiliation{Department of Electrical and Electronic Engineering, University of Dhaka, Dhaka-1000, Bangladesh}

\author{Shovon Chakraborty}
\affiliation{Department of Electrical and Electronic Engineering, University of Dhaka, Dhaka-1000, Bangladesh}

\author{Imtiaz Ahmed}\thanks{imtiaz@du.ac.bd}
\affiliation{Department of Electrical and Electronic Engineering, University of Dhaka, Dhaka-1000, Bangladesh}

\begin{abstract}
We perform first principles simulations for the structural, elastic and electronic properties of orthorhombic samarium orthoferrite SmFeO\textsubscript{3} within the framework of density functional theory. A number of different density functionals, such as local density approximation, generalized gradient approximation, Hubbard interaction modified functional, modified Becke-Johnson approximation and Heyd–Scuseria–Ernzerhof hybrid functional have been used to model the exact electron exchange-correlation. We estimate the energy of the ground state for different magnetic configurations of SmFeO\textsubscript{3}. The crystal structure of SmFeO\textsubscript{3} is characterized in terms of the lattice parameters, atomic positions, relevant ionic radii, bond lengths and bond angles. The stability of the SmFeO\textsubscript{3} orthorhombic structure is simulated in terms of its elastic properties. For the  electronic structure simulations, we provide estimates based on density functionals with varying degrees of computational complexities in the Jacob's ladder.  
 
 \vspace{5mm}

\end{abstract}

\
\maketitle

\section{\label{sec:level2}Introduction}
The rare-earth orthoferrites managed to reign in the active field of materials research for more than half a century \cite{white1969review}. These materials have common chemical formula $R$FeO\textsubscript{3} where $R$ is a rare-earth ion in the lanthanide series. Originally $R$FeO\textsubscript{3} materials were studied as a family of canted anti-ferromagnets which revealed exciting, novel; sometimes baffling magnetic properties \cite{bozorth1958origin, treves1962magnetic, yamaguchi1973magnetic}.
One of the prominent members of the rare-earth orhtoferrites is the samarium orthoferrite SmFeO\textsubscript{3} (SFO hereafter). SFO is a promising candidate for many spintronic device applications for many of its interesting and intriguing properties; such as spontaneous reversal of at cryogenic temperatures below $4$ K, fast magnetic switching capabilities with high spin switching temperature of $278.5$ K and high spin axis rotation temperature of $480$ K \cite{cao2014temperature}. SFO magnetic properties depend on particle size,  surface morphology and measurements temperature  \cite{selvadurai2019metamagnetism, chaturvedi2017nanosize, ahlawat2018influence, wang2016room, fu2015ultralow}. Its high magnetostriction coefficient along with the anomalous magneto-electric behaviour may open up possibilities for different magnetoelastic devices \cite{abe1977magnetostriction, zhou2014hydrothermal, ahlawat2016tunable}.

SFO has also found its applications in high performance electrode materials for solid state lithium-ion batteries \cite{liu2018smfeo3}, as good dielectric materials for electronics \cite{prasad2011abnormal, khan2018magneto}, in photocatalytic applications for renewable energy technology \cite{tang2016microwave, maity2019investigation} and also in multiferroics \cite{zhang2016multiferroicity, liu2020structure}. A material with such diverse applications also embodies rich physics due to its bewildering exchange interaction between the $4f$ electrons in the rare-earth Sm and $3d$ electrons in transition element Fe. This prompted the need for understanding this fascinating material from quantum mechanical first principles calculations within the framework of density functional theory (DFT) \cite{jones2015density, zhao2016origin}. The DFT based simulations in combination with experimental observations opened up fascinating debate about the origin (complex interplay between the inverse Dzyaloshinskii-Moriya and exchange-striction) and existence of ferroelectric ordering in SFO \cite{lee2011spin, johnson2012comment, lee2012lee, kuo2014k}. The SFO vibrational phonon frequency and associated Raman modes have been studied using DFT simulations along with experimental investigations \cite{weber2016raman,tyagi2018detail, bhadram2013spin, gupta2018temperature}. The ab initio DFT based simulations have been performed to understand electronic structure of a number of $R$FeO\textsubscript{3} materials including SFO, where the presence of $4f$ electrons in Sm caused diffcuties in predicting correct electronic ground state configuration \cite{iglesias2005ab, singh2008electronic}. An improved electronic band structure calculation of SFO based on incorporating Hubbard interaction gave a better estimates for the experimentally measured electronic properties of SFO \cite{triguk2016electronic}.

Although numerous experimental work on SFO exist in literature, to the best of our knowledge, the DFT based first principles calculations to investigate its structural, elastic and electronic properties in a systematic manner are hard to find in existing studies. Here we perform DFT simulations for structural, elastic and electronic properties of SFO in an organized and methodical fashion. We explore different density functionals with varying degrees of computational complexities within the DFT framework to make a comparative analysis and study their potentials in explaining the relevant physical properties of SFO.

\section{\label{sec:level2}Computational Details}
We perform DFT based spin-polarized and
non spin-polarized simulations within the framework projector augmented wave (PAW) method using the Vienna $Ab$ $Initio$ Simulation Package (VASP) \cite{kresse1996efficient, kresse1999ultrasoft}. We consider a  SFO unit cell consists of four Sm atoms, four Fe atoms and twelve O atoms;  a total of 20 atoms are considered for all simulations performed in this paper. For the PAW, we divide the SFO electron configuration into core and valence categories. We considered sixteen electrons of Sm ($4f^55s^25p^65d^16s^2$), 8 electrons of Fe ($3d^64s^2$) and six electrons of O ($2s^22p^4$) as valence electrons (in total thirty valence electrons) and the remaining electrons are treated within the frozen core approximation. Structural relaxation and optimization are carried out by sampling the Brillouin zone with  a $5\times5\times3$ Monkhorst Pack grid k-points mesh until the Hellmann–Feynman forces reached 0.005 eV/$\text{\AA}$. We used the self-consistent total energy convergence of $10^{-8}$ eV.  For truncating the plane wave expansion for the PAW, a plane wave energy cutoff of 480 eV is used in all simulations; except for the case of elastic properties where an increased energy cutoff of 520 eV is used to ensure convergence.

We used a number of different approximations for the unknown exchange-correlation term in the Kohn-Sham Hamiltonian \cite{kohn1965self}. We use Ceperley-Alder local density approximation (LDA) where the exchange term is obtained from the homogeneous electron gas and the correlation term is approximated from numerically accurate Monte Carlo methods \cite{ceperley1980ground}. The semi local generalized gradient approximation (GGA) is implemented with three different standard variants Perdew-Wang (PW91) \cite{perdew1992atoms}, Perdew-Burke-Ernzerhof (PBE) \cite{perdew1996generalized} and its optimized version PBEsol \cite{perdew2008restoring}. We also make use of the “Hubbard-U” scheme for LDA and GGA-PBE which are referred to as LDA+U and  GGA-PBE+U \cite{dudarev1998electron, anisimov1991band}. To boost the diluted Coulomb interaction for the localized orbitals, we used U = 6 eV for Sm and U = 4 eV for Fe atoms. These are common choices for the on site Coulomb interaction term U in case of  rare-earth and transition metal atoms that produce correct materials properties \cite{dieguez2011first, zhao2016origin, stroppa2010multiferroic, triguk2016electronic}. We also explore the modified Becke and Johnson (mBJ) exchange potential in combination with GGA-PBE in case of electronic structure calculation \cite{tran2009accurate}. For more accurate estimations for electronic structure, computationally intense Heyd–Scuseria–Ernzerhof (HSE06) hybrid functional based simulations have been performed \cite{heyd2003hybrid,krukau2006influence, paier2006screened}.

\begin{figure}
	\begin{center}
		\includegraphics[scale=1]{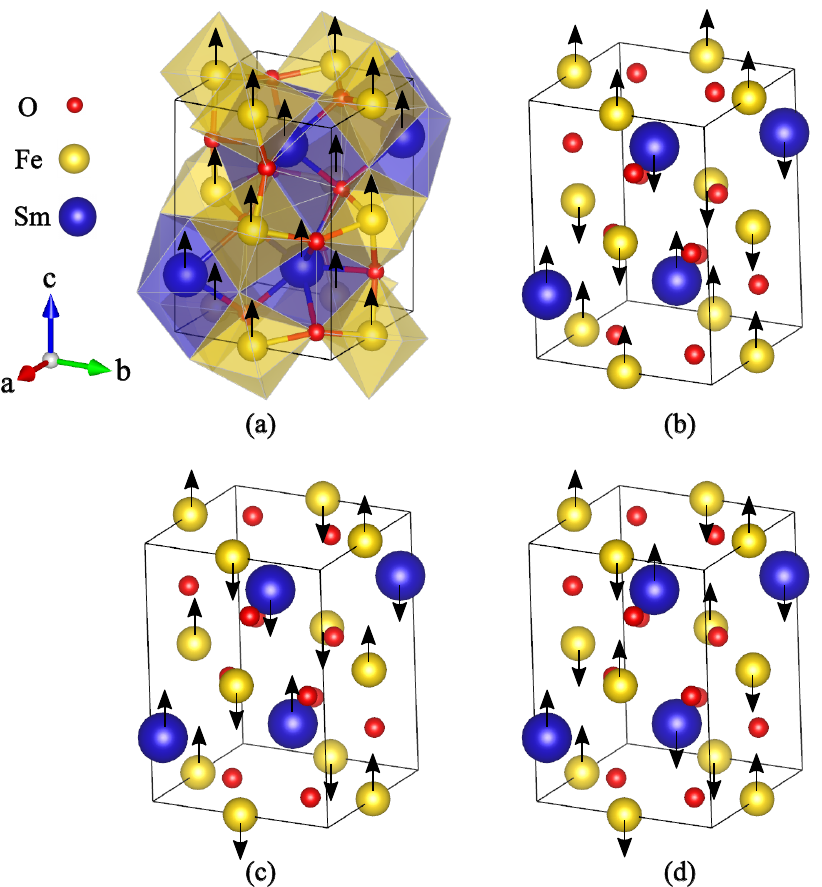}
		\caption{ The SFO unit cell with four different spin magnetic  configurations, (a) Ferromagnetic (FM), (b) C-type antiferromagnetic (C-AFM), (c) A-type antiferromagnetic (A-AFM) and (d)  G-type antiferromagnetic (G-AFM). Black arrow indicates the spin magnetic moment orientations of atoms.}
		\label{Figure_00}
	\end{center}
\end{figure}

The SFO can exist in four different magnetic structures; one ferromagnetic (FM) and three antiferromagnetic (AFM) which are A-AFM, C-AFM and G-AFM, see Fig.~\ref{Figure_00}(a-d). We have calculated total energies for all four magnetic configurations for LDA+U, GGA-PBE and GGA-PBE+U functionals in the the Kohn-Sham Hamiltonian. In all cases, G-AFM turned out to be the magnetic configuration with minimum total energy which is consistent with results to be found in \cite{iglesias2005ab,singh2008electronic, kuo2014k}. Assuming the G-AFM as our minimum energy reference, we estimated the energy differences $\Delta E_\text{FM-GAFM}$, $\Delta E_\text{AAFM-GAFM}$ and $\Delta E_\text{CAFM-GAFM}$  for the three higher energy states FM, A-AFM and C-AFM respectively, see Table~\ref{table:E_diff}. All spin-polarized calculations presented in this article are for G-AFM magnetic configuration in SFO.
 
\begin{table}[h]
\centering
\begin{tabular}{c c c c }
\hline
Method & LDA+U & GGA-PBE & GGA-PBE+U\\
\hline
$\Delta E_\text{FM-GAFM}~(\text{eV})$ & 1.219 & 1.063 & 0.828  \\
$\Delta E_\text{AAFM-GAFM}~(\text{eV})$ &  1.5 & 7.001 & 0.575\\
$\Delta E_\text{CAFM-GAFM}~(\text{eV})$ & 0.726 & 0.13 & 0.220 \\
\hline
 \end{tabular}
 \caption{\label{table:E_diff}Energy differences between different SFO magnetic configurations calculated for LDA+U, GGA-PBE and GGA-PBE+U.}
 \end{table}

\section{\label{sec:level2}Crystal Structure}
\begin{table*}[t]
\centering
\begin{tabular}{ |c|c|c|c|c|c|c|c|c|c|c|}
 \hline
 \multicolumn{11}{|c|}{Lattice Parameters} \\
 \hline
 
 Methods& \multicolumn{5}{|c|}{Spin-Polarized }& \multicolumn{5}{|c|}{Non Spin-Polarized }\\
 \hline
  &a (\AA) & b (\AA) & c (\AA) &V($\text{\AA}^3$)& $\alpha=\beta=\gamma$ (deg.) & a (\AA) &b (\AA) & c (\AA) & V($\text{\AA}^3$)&$\alpha=\beta=\gamma$ (deg.)\\
  \hline 
 LDA & 5.203 & 5.277 & 7.392&202.96 & 90 & 4.972 & 5.484 & 7.23 8 &197.35 & 90\\
 \hline
 LDA+U & 5.301 & 5.529 & 7.586 &222.34 & 90 & 5.134 & 5.448 & 7.298 & 204.12 & 90\\
 
 \hline
 GGA-PW91 & 2.574 & 10.953 & 9.443 & 266.23 & 90 & 7.543 & 4.574 & 8.876&306.23 & 90\\
 \hline
 GGA-PBE & 5.436 & 5.648 & 7.654 & 234.99 & 90 & 5.125  & 5.590 & 7.397& 211.91 & 90\\
 \hline
 GGA-PBE+U & 5.428 & 5.657 &  7.761 & 238.31 & 90 & 5.317 & 5.540 & 7.516 & 221.39 & 90\\
 \hline
 GGA-PBEsol & 5.258 & 5.285 & 7.438& 206.69& 90 & 5.063 & 5.579 & 7.185&202.95 & 90\\
 \hline
 Exp.&\multicolumn{2}{|c|}{a= 5.39 \AA}&\multicolumn{2}{|c|}{b=5.58 \AA }&\multicolumn{2}{|c|}{c=7.71 \AA }&\multicolumn{2}{|c|}{V = 231.89 $\text{\AA}^3$ }& \multicolumn{2}{|c|}{$\alpha=\beta=\gamma$=90}\\
 \hline
  \end{tabular}
  \caption{\label{table:Lattice_Param} Structural lattice parameter of SFO calculated from LDA, LDA+U, GGA-PW91, GGA-PBE, GGA-PBE+U and GGA-PBEsol for both spin-polarized and non spin-polarized configurations. The experimental (Exp.) lattice parameter values can be found in \cite{treves1962magnetic,liu2018smfeo3, maity2019investigation, gupta2018temperature}. }
  \label{table:1}
\end{table*}
\begin{table*}
\centering
\begin{tabular}{ |c|c|c|c|c|c|c|c|c|c|c|}
 \hline
 \multicolumn{11}{|c|}{Atomic Positions in Wyckoff Coordinate} \\
 \hline
 \multicolumn{2}{|c|}{} &\multicolumn{3}{|c|}{LDA+U}&\multicolumn{3}{|c|}{GGA-PBE}&\multicolumn{3}{|c|}{GGA-PBE+U}\\
 \hline
 Atom & Site & x & y & z & x & y & z& x & y & z\\
 \hline
  Sm & 4a &  0.9854 & 0.0592 & 0.2507 & 0.9897
& 0.0548 & 0.2506 & 0.9852 & 0.0591 &  0.2503\\
  \hline
  Fe & 4b &  0.0000 & 0.4966 & 0.0000 & 0.0000 & 0.4966 & 0.0005 & 0.9999 & 0.4994 & 0.0000\\
  \hline
  O1 & 8d & 0.7001 & 0.2982 & 0.0487 & 0.6982 & 0.3049 & 0.0452 & 0.6997 & 0.2994 & 0.0503\\
  \hline
  O2 & 4c & 0.0930 & 0.4735 & 0.2498 & 0.0902 & 0.4735 & 0.2499 & 0.0959 & 0.4696 & 0.2499\\
  \hline
  \end{tabular}
  \caption{\label{table:Wyckoff_Pos} Atomic Positions Sm, Fe and O atoms in SFO unit cell in terms of Wyckoff coordinates calculated from  LDA+U, GGA-PBE and GGA-PBE+U for G-AFM spin-polarized configuration.}
  \label{table:1}
\end{table*}
\begin{table*}[]
    \centering
    \begin{tabular}{|c|c|c|c|c|c|c|c|c|c|}
     \hline
         Parameter & \multicolumn{3}{|c|}{LDA+U}& \multicolumn{3}{|c|}{GGA-PBE} & \multicolumn{3}{|c|}{GGA-PBE+U}\\
         \hline
         Ionic Radius (\AA) & \multicolumn{3}{|c|}{Sm= 1.482, Fe = 1.302, O = 0.8} & \multicolumn{3}{|c|}{Sm = 1.482, Fe = 1.302, O = 0.82}& \multicolumn{3}{|c|}{Sm = 1.482, Fe = 1.302, O = 0.82}\\
         $d_\text{Sm-O}$ (\AA) & \multicolumn{3}{|c|}{2.28, 2.51, 2.65} & \multicolumn{3}{|c|}{2.33, 2.43, 2.65} & \multicolumn{3}{|c|}{ 2.33, 2.40, 2.73}\\
         $d_\text{Fe-O}$ (\AA) & \multicolumn{3}{|c|}{1.96, 1.980, 2.00} & \multicolumn{3}{|c|}{1.98, 2.0, 2.1} & \multicolumn{3}{|c|}{ 2.015, 2.027, 2.051}\\
        $d_\text{Sm-Fe}$ (\AA) & \multicolumn{3}{|c|}{3.088, 3.21, 3.341} & \multicolumn{3}{|c|}{3.15, 3.30, 3.38,  3.66}& \multicolumn{3}{|c|}{ 3.16, 3.29,3.71}\\
        $\Theta_\text{Fe-O-Sm} (deg.)$ & \multicolumn{3}{|c|}{85.76, 86.03, 90.21} & \multicolumn{3}{|c|}{87.77, 88.82, 90.62}& \multicolumn{3}{|c|}{85.789, 89.785, 89.953}\\
        $\Theta_\text{Fe-O-Fe}$ (deg.) & \multicolumn{3}{|c|}{149.39} &\multicolumn{3}{|c|}{148.61, 149.61, 150.00} & \multicolumn{3}{|c|}{148.433, 148.440, 148.597}\\
        $\Theta_\text{O-Fe-O}$ (deg.) & \multicolumn{3}{|c|}{88.57, 89.27, 90.86} & \multicolumn{3}{|c|}{87.70, 88.92, 90.43}& \multicolumn{3}{|c|}{88.41, 89.797, 90.012}\\
        \hline
        
    \end{tabular}
    \caption{\label{table:Bond_Length} Calculated ionic radii, Sm-O, Fe-O, Sm-Fe bond lengths, Fe-O-Sm, Fe-O-Fe and O-Fe-O bond angles for spin polarized  LDA+U, GGA-PBE and GGA-PBE+U.}
\end{table*}

The SFO has a distorted orthorhombic perovskites unit cell structure with the space group \textit{Pnma} (no. 62) at room temperature where the magnetic easy axis is along c axis of the orthorombic unit cell \cite{hahn1997symmetry, singh2008electronic, iglesias2005ab, kuo2014k, maity2019investigation}. Each unit cell ($\text{a}<\text{b}<\text{c}$) of SFO has four $\text{Sm}^{3+}$ ions at the centers and four $\text{Fe}^{3+}$ at the corners surrounded by oxygen octahedra which are tilted along the crystallographic b-axis, see Fig.~\ref{Figure_00}(a). The orthorhombic structural distortion is due to the size mismatch between the octahedral holes available for the $\text{Sm}^{3+}$ in the unit cell and the actual smaller $\text{Sm}^{3+}$ ion.

The SFO has experimentally measured mutually orthogonal ($\alpha=\beta=\gamma=$90 deg) lattice constants a $= 5.39$ $\text{\AA}$, b $= 5.58$ $\text{\AA}$, c $= 7.71$ $\text{\AA}$ with a unit cell volume V= 231.89 $\text{\AA}^3$ \cite{treves1962magnetic,liu2018smfeo3, maity2019investigation, gupta2018temperature}. We performed structural optimization of the unit cell with different variants of LDA and GGA in both spin-polarized and non spin-polarized configurations. The simulated values for lattice parameters a, b, c, V, $\alpha$, $\beta$ and $\gamma$ are summarized in Table~\ref{table:Lattice_Param}. For non spin-polarized calculations, all variants of LDA and GGA show significant deviations resulting in poor estimates for lattice parameters as compared to experimental values. Lattice parameters calculated from GGA-PW91 functional produced inconsistent lattice parameters even in the case of spin-polarized calculations. The basic LDA provides more reasonable estimates for lattice parameters in the spin-polarized configuration. Moreover  LDA+U, GGA-PBE and GGA-PBE+U also provide good estimation for lattice parameters in case of spin-polarized calculations. Both GGA-PBE and GGA-PBE+U consistently overestimated the unit cell volume V by 1.34 \% and 2.77 \% respectively whereas LDA+U under estimate it by 4.12 \%. This is consistent with the fact that LDA usually over binds the atoms inside the unit cell and the semi local GGA does the opposite. Based on these reliable estimates, we calculate the atomic positions of four $\text{Sm}^{3+}$, four $\text{Fe}^{3+}$ and twelve O atoms in the unit cell using Wyckoff coordinates \cite{hahn1997symmetry,zhu2017electronic}, see Table~\ref{table:Wyckoff_Pos}. For example after internal GGA-PBE structure relaxation, the Sm atom occupies the (0.9897, 0.0548, 0.2506) site, the Fe atom in (0.0000, 0.4966, 0.0005) site, the O1 atom in the (0.6982, 0.3049, 0.0452) site, and the O2 atom in (0.0902, 0.4735, 0.2499) sites in Wyckoff coordinates. We calculate  ionic radius, bond lengths and bond-angles relevant for SFO, see Table~\ref{table:Bond_Length} \cite{chaturvedi2016dynamics, maity2019investigation, zhang2010systematic}. Three GGA-PBE optimized Sm–O bond lengths are 2.33, 2.43, 2.65 $\text{\AA}$ (in case of LDA+U, 2.28, 2.51, 2.65 $\text{\AA}$ ) are slightly larger than the Fe–O bond lengths of  1.98, 2.0, 2.1 $\text{\AA}$ (in case of LDA+U, 1.96, 1.980, 2.00 $\text{\AA}$). These are in line with the relation of ionic radii, $\text{Sm}^{3+}>\text{Fe}^{3+}$. The distances between $\text{Sm}^{3+}$ and $\text{F}e^{3+}$ ions are 3.136, 3.284 and 3.362 $\text{\AA}$  for GGA-PBE (3.088, 3.21, 3.341 $\text{\AA}$ in case of LDA+U). The bond angles of Fe–O–Sm are 87.77$^{\circ}$, 88.82$^{\circ}$, 90.62$^{\circ}$ in GGA-PBE relaxed structure (85.76$^{\circ}$, 86.03$^{\circ}$, 90.21$^{\circ}$ for LDA+U), deviating from the ideal value of 90$^{\circ}$. This indicates the presence of structural distortion in the relaxed SFO unit cell.

\section{\label{sec:level2}Elastic Properties}

\begin{table}[h!]
\centering
\begin{tabular}{c c c c }
\hline
Elastic  & \multicolumn{3}{c}{Spin-Polarized} \\
Properties & \\
\hline
& LDA & LDA+U & GGA-PBE\\
\hline
$C_\textsubscript{11}$ (GPa) & 211.923 & 288.021 & 181.250\\
$C_\textsubscript{12}$(GPa) & 63.803 & 96.286 & 96.419 \\
$C_\textsubscript{13}$(GPa) & 58.391 & 151.939 & 75.115\\
$C_\textsubscript{22}(GPa)$ & 143.286 & 215.376 & 223.149\\
$C_\textsubscript{23}$(GPa) & 25.618 & 132.203 & 121.135\\
$C_\textsubscript{33}$(GPa) & 237.943 & 293.781 & 199.397\\
$C_\textsubscript{44}$(GPa) & 98.739 & 96.396 & 84.671\\
$C_\textsubscript{55}$(GPa) & 85.035 & 82.377 & 68.517\\
$C_\textsubscript{66}$(GPa) & 93.821 & 44.923 & 35.920\\
$B_\textsubscript{V}$(GPa) & 98.753 & 173.115 & 132.126\\
$B_\textsubscript{R}$(GPa) & 93.110 &  164.829 & 127.968\\
$B_\textsubscript{H}$(GPa) & 95.931 & 168.972 & 130.047\\
$G_\textsubscript{V}$(GPa) & 85.208 & 72.523 & 58.563\\
$G_\textsubscript{R}$(GPa) & 80.554 & 66.73 & 53.057\\
$G_\textsubscript{H}$(GPa) & 82.881 & 69.627 & 55.810\\
$E_\textsubscript{V}$(GPa) & 198.526 & 190.909 & 153.074\\
$E_\textsubscript{R}$(GPa) & 187.570 & 176.392 & 139.844\\
$E_\textsubscript{H}$(GPa) & 193.048 & 183.656 & 146.477 \\
$\nu_\textsubscript{V}$ & 0.165 & 0.316 & 0.307\\
$\nu_\textsubscript{R}$ & 0.164 & 0.322 & 0.318\\
$\nu_\textsubscript{H}$ & 0.165 & 0.319 & 0.312\\
\hline
 \end{tabular}
 \caption{\label{table:Elastic_Cons} Elastic constants ($C_\textsubscript{ij}$), bulk modulus ($B_\textsubscript{V}$, $V_\textsubscript{R}$ and $B_\textsubscript{H}$), shear modulus ($G_\textsubscript{V}$, $G_\textsubscript{R}$, $G_\textsubscript{H}$), Young's modulus ($E_\textsubscript{V}$, $E_\textsubscript{R}$, $E_\textsubscript{H}$), Poisson ratio ($\nu_\textsubscript{V}$, $\nu_\textsubscript{R}$ and $\nu_\textsubscript{H}$ ) in Voigt–Reuss–Hill framework for G-AFM orthorhombic SFO using LDA, LDA+U and GGA-PBE.} 

 \end{table}
 
To investigate the structural stability of the G-AFM orthorhombic SFO, we calculated elastic tensor $C_\textsubscript{ij}$ by applying forces thereby creating six finite perturbation to the lattice and measuring the $C_\textsubscript{ij}$ from the standard strain-stress relationship \cite{wu2005systematic, shang2007first}. To ensure the convergence of the stress tensor we use the plane wave energy cutoff to be 520 eV for the PAW.  For orthorhombic SFO, we have six non-zero independent elastic constants $C_\textsubscript{11}$, $C_\textsubscript{12}$, $C_\textsubscript{13}$, $C_\textsubscript{22}$, $C_\textsubscript{23}$, $C_\textsubscript{33}$, $C_\textsubscript{44}$, $C_\textsubscript{55}$ and $C_\textsubscript{66}$. From these non-zero $C_\textsubscript{ij}$s, we can check the necessary and sufficient Born criteria for mechanical stability for an orthorhombic system
\begin{equation}
    C_\textsubscript{11}>0, \:C_\textsubscript{44}>0, \:C_\textsubscript{55}>0, \:C_\textsubscript{11}C_\textsubscript{22}>C_\textsubscript{12}^2
\end{equation}
\begin{equation}
C_\textsubscript{11}C_\textsubscript{22}C_\textsubscript{33}+ 2C_\textsubscript{12}C_\textsubscript{13}C_\textsubscript{23}-C_\textsubscript{11}C_\textsubscript{23}^2-C_\textsubscript{22}C_\textsubscript{13}^2-C_\textsubscript{33}C_\textsubscript{12}^2>0,
\end{equation}
are satisfied in all three cases, i.e., LDA, LDA+U and GGA-PBE+U \cite{mouhat2014necessary}, see Table~\ref{table:Elastic_Cons}. The other important elastic properties like bulk-modulus (B), shear modulus (G), Young's modulus (E), Poisson's ratio ($\nu$) are calculated using three different theories such as Reuss ($B_\textsubscript{R}$, $G_\textsubscript{R}$, $E_\textsubscript{R}$ and $\nu_\textsubscript{R}$), Voigt ($B_\textsubscript{V}$, $G_\textsubscript{V}$, $E_\textsubscript{V}$ and $\nu_\textsubscript{V}$) and Hill ($B_\textsubscript{H}$, $G_\textsubscript{H}$, $E_\textsubscript{H}$ and $\nu_\textsubscript{H}$) \cite{hill1952elastic, dong2013elastic, yaakob2013first}. We note that the values for Voigt–Reuss–Hill bulk moduli are different for the orthorhombic for unit cell indicating the departure from the cubic symmetry (they are equal only for unit cell with Cubic symmetry). The LDA predicts the highest values for shear modulus ($G_\textsubscript{H} = 82,881$ GPa) and Young's modulus $E_\textsubscript{H} = 193.048$ GPa) as compared to LDA+U and GGA-PPE. This implies both resistance to plastic deformation and stiffness of SFO are largest within the LDA exchange-correlation framework. This is indicative of atoms in SFO unit cell being over bounded in LDA. The Hubbard U term in case of LDA+U corrects for the binding energies of atoms in SFO unit cell which results in reduction of both $G_\textsubscript{H}$ and $E_\textsubscript{H}$ to $69.627$ GPa and $183.656$ GPa respectively. In case of GGA-PBE, the values for $G_\textsubscript{H}$ and $E_\textsubscript{H}$ reduces even further to $55.810$ GPa and $146.477$ GPa bearing the signature of under bounded atoms in the unit cell. For ductile/brittle test, the estimated Pugh ratio for LDA, LDA+U and GGA-PBE are $0.864$, $0.412$ and $0.429$ respectively; all of which are smaller than the critical value of $1.75$ indicating the brittle nature  of orthorhombic SFO \cite{pugh1954xcii}. This is corroborated with the estimated values of Poisson's ratio ($\nu_\textsubscript{V}$, $\nu_\textsubscript{R}$ and $\nu_\textsubscript{H}$) being smaller than the critical value of 0.33 in the Voigt–Reuss–Hill framework.

\section{\label{sec:level2}Electronic Structure}
\begin{figure}
	\begin{center}
		\includegraphics[scale=1]{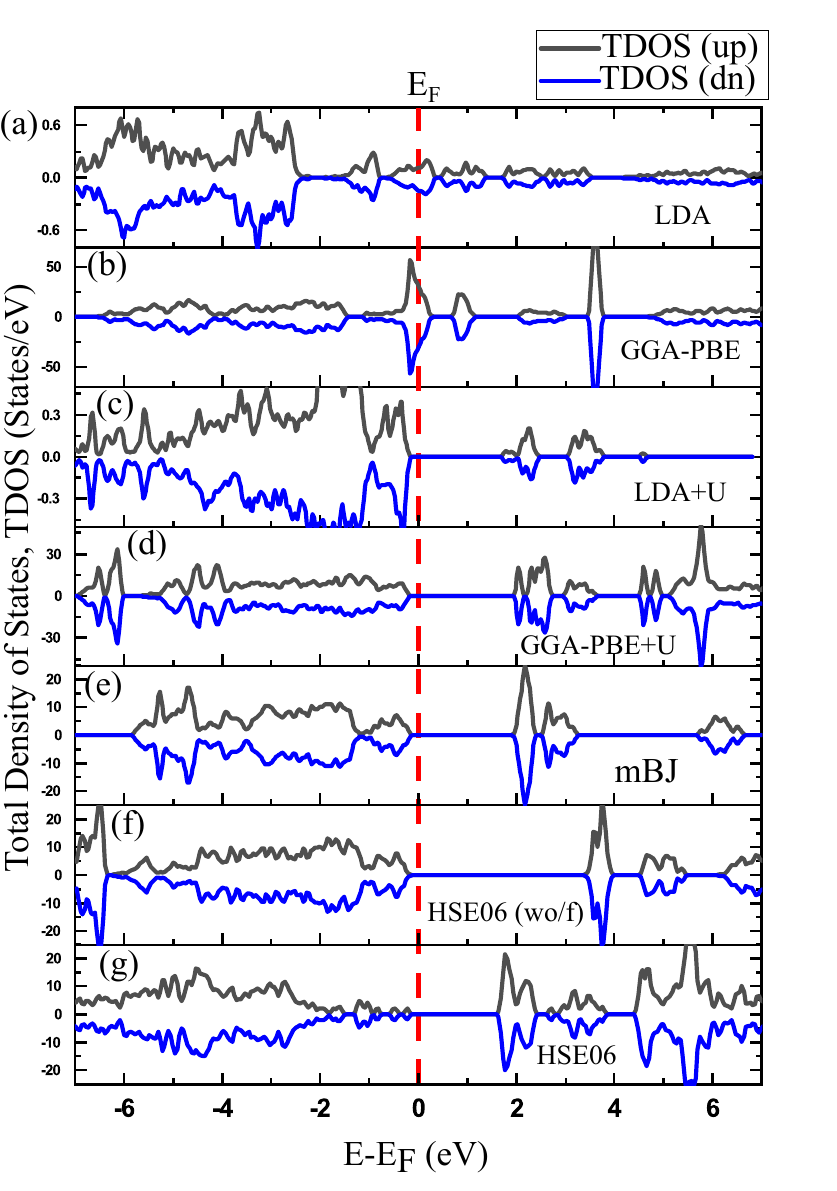}
		\caption{ Total density of states TDOS of orthorhombic SFO calculated with (a) LDA, (b) GGA-PBE, (c) LDA+U, (d) GGA-PBE+U, (e) mBJ, (f) HSE06 with Sm-$4f$ orbital in core, (g) HSE06 with Sm-$4f$ orbital in valence for PAW method. Due to AFM ordering, symmetry exists between the spin-up in the upper part and and spin-down in the lower part of the TDOS. The red dashed line indicates the position of the Fermi level $E_F$.}
		\label{Figure_01}
	\end{center}
	
\end{figure}
\begin{figure}
	\begin{center}
		\includegraphics[scale=1]{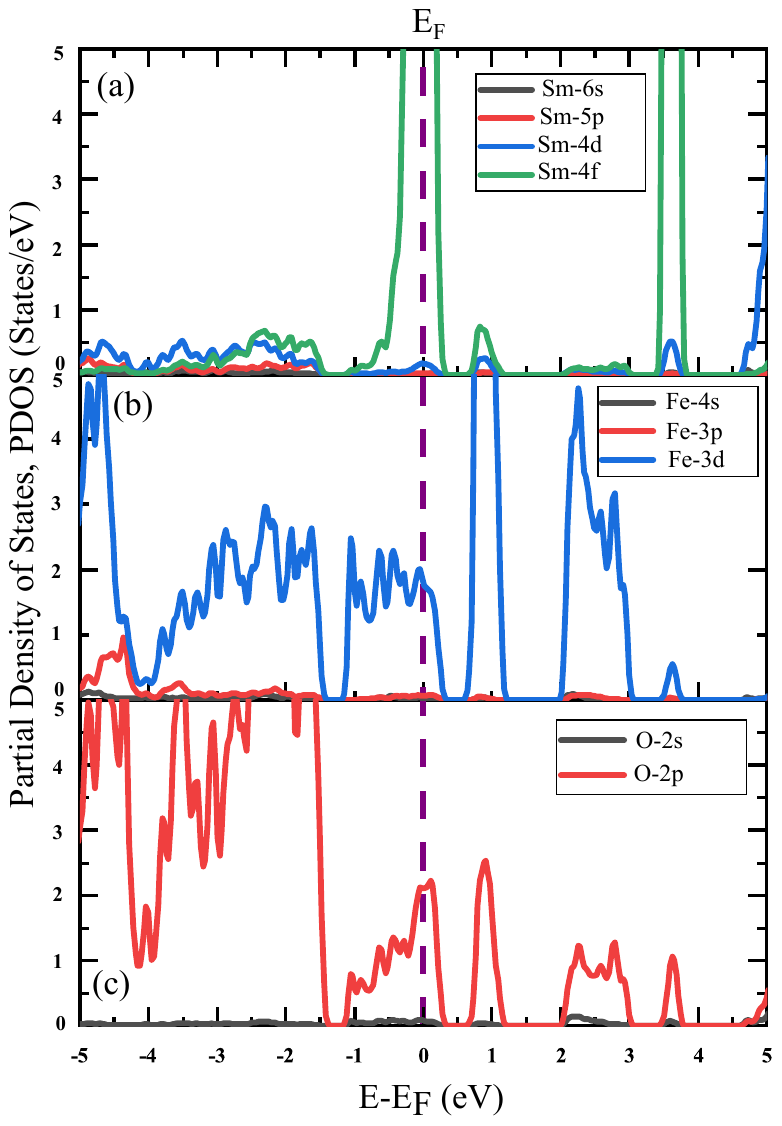}
		\caption{Partial density of states PDOS by projecting TDOS on to (a) Sm-$6s$, Sm-$5p$, Sm-$4d$, Sm-$4f$, (b) Fe-$4s$, Fe-$3p$, Fe-$3d$, (c) O-$2s$, O-$2p$ for GGA-PBE. Due to the AFM spin symmetry considerations, only PDOS for up spin channel is plotted for clarity.}
		\label{Figure_02}
	\end{center}
	
\end{figure}

\begin{figure}
	\begin{center}
		\includegraphics[scale=1]{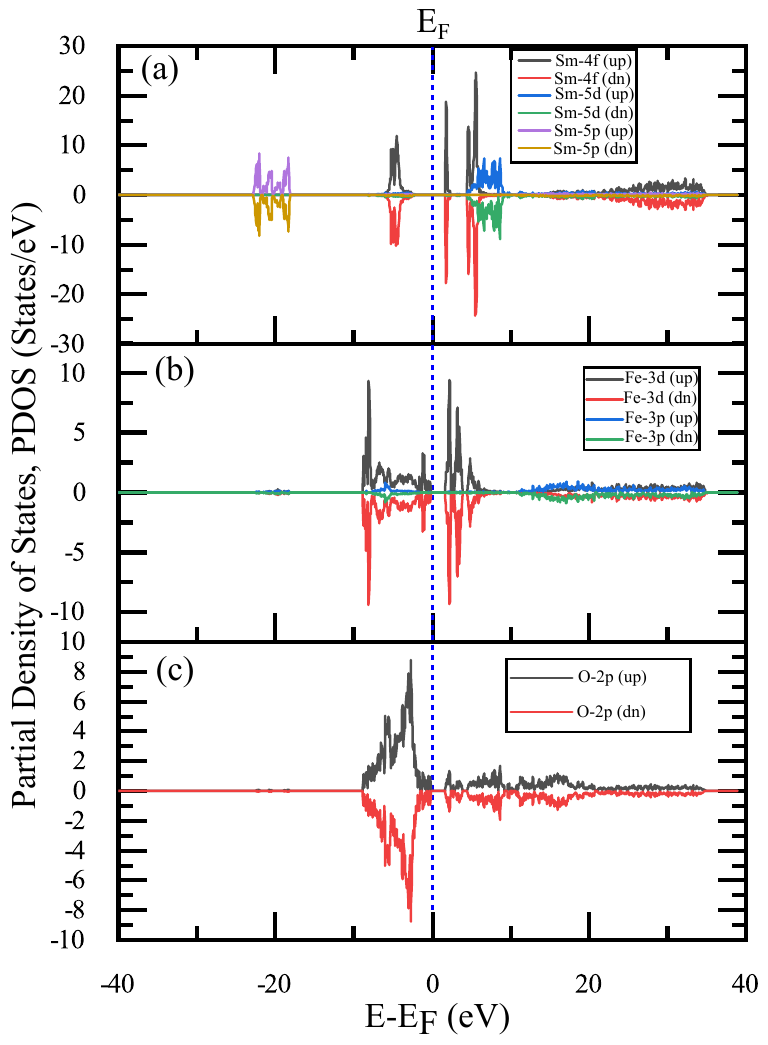}
		\caption{Spin resolved PDOS by projecting onto (a) Sm-$4f$, Sm-$5d$, Sm-$5p$ and Sm-$4f$, (b) Fe-$3p$ and Fe-$3d$, (c) O-$2s$ and O-$2p$ for HSE06 hybrid functional with Sm-$4f$ electrons treated as valence in PAW method.}
		\label{Figure_03}
	\end{center}
	
\end{figure}
\label{Figure_03}
\begin{figure}
	\begin{center}
		\includegraphics[scale=1]{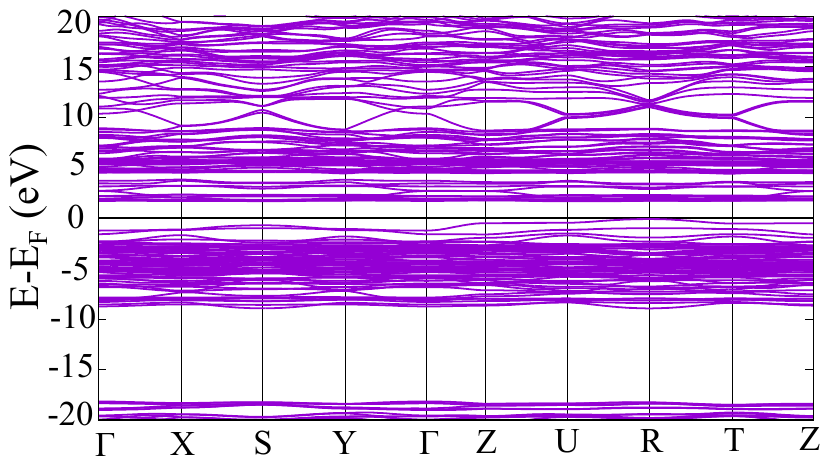}
		\caption{Electronic band structure along high symmetry k-points $\Gamma$, R, S, T, U, Y and Z in the orthorhombic SFO Brillouin zone for HSE06 hybrid functional.}
		\label{Figure_04}
	\end{center}
	
\end{figure}
To analyze the electronic structure of SFO, we calculate the spin-resolved total density of states (TDOS) as a function of energy with a $14$ eV energy window centered at the Fermi level ($E_F$) for different exchange-correlation functionals, see Fig.~\ref{Figure_01}. Due to the AFM ordering, the density of states are equal for spin up and down configuration in all cases. 
Although the LDA and semi-local GGA-PBE are computationally cheaper in comparison with more sophisticated methods, they result in non-zero TDOS at $E_F$ indicating a metallic behavior for G-AFM SFO, see Fig.~\ref{Figure_01}(a, b). In case of LDA similar metallic behaviour for SFO can be found in \cite{iglesias2005ab, singh2008electronic}. But this metallic state of SFO is inconsistent with experimentally measured electrically resistive  nature of orthorhombic $Pnma$ SFO upto the Neel temperature $T_N = 670$ K \cite{maslen1996synchrotron, lee2011spin}. This discrepancy in case of LDA and GGA-PBE can be attributed to inadequate description of strong Coulomb repulsion between the electrons in localized partially filled $d$ (in Fe) and $f$ orbitals (in Sm) in SFO. The on site Hubbard U interaction term in case of LDA+U opens up a gap of 1.86 eV between the highest occupied (highest occupied molecular orbital, HOMO) and lowest unoccupied (lowest unoccupied molecular orbital, LUMO) energies in TDOS at the Fermi level; see Fig.~\ref{Figure_01}(c)  which is consistent with the results obtained from DFT simulation using a different ABINIT software package \cite{triguk2016electronic, gonze2005brief}. In case of the semi local GGA-PBE+U, the insulating energy gap is found to be 2 eV in Fig.~\ref{Figure_01}(d). The Hubbard U term for both LDA+U and GGA-PBE+U is a semiempirical parameter that needs to be optimized depending on the type of materials. The choice of material dependent U parameter is usually adhoc in nature. A more systematic and rigorous method, such as mBJ, can be found by climbing one step in the Jacob's ladder with increased computational complexity \cite{perdew2001density, perdew2010fourteen}. The mBJ method which is usually used in combination with the LDA where the LDA exchange potential is replaced by the mBJ potential leaving the electron correlation potential unchanged \cite{zhu2017electronic}. Here we implement the mBJ on top of GGA-PBE, i.e. the GGA-PBE excahnge is treated with mBJ leaving the electron correlation unchanged. The mBJ exchange potential has outperformed basic LDA and GGA-PBE in electronic structure calculations and provides more accurate DOS and bandgap for different semiconducting and insulating materials \cite{singh2010optical, koller2012improving, fan2013half}. In our calculation this is evident from the fact that mBJ resulted in 2 eV energy gap in TDOS around $E_F$ similar to the GGA-PBE+U methods, see see Fig.~\ref{Figure_01}(e). Now we climb one more step in the Jacob's rung and implement the HSE06 hybrid functional in which only the exchange interaction of the GGA-PBE is divided into short and long range parts leaving the electron correlation part unchanged. The 25\% of the short range GGA-PBE exchange is replaced by the exact Hartree–Fock exchange with screening parameter of 0.2 ${\AA}^{-1}$ for the inter electronic Coulomb potential. Now within the HSE06 framework, treating the Sm-$4f$ electrons as valence imposes significant computational complexities in structural relaxation and self-consistent energy calculations. At first we treat the Sm-$4f$ electron as a core to keep the computational complexities and convergence issues manageable and found a 3.42 eV gap around the $E_F$ in TDOS which significantly higher all other methods mentioned above, see Fig.~\ref{Figure_01}(f). Although the phonon frequency and density of states in rare earth ortho ferrites are shown to have small effect  whether the Sm-$4f$ electrons are treated as core or valence \cite{weber2016raman}; electronic density of states are expected to depend on it. Hence we implement the HSE06 treating the Sm-$4f$ electrons as valence with higher computation cost and obtained 1.73 eV energy gap, see Fig.~\ref{Figure_01}(g); which is much smaller compared to the case when  Sm-$4f$ electrons are treated as core in PAW method above.

To analyze the structure of the density of states, we projected the TDOS onto the atomic orbitals of individual ions in SFO and calculate the partial density of states (PDOS) of Sm, Fe and O. First we present the PDOS for the semilocal GGA-PBE to show explicitly  why it fails to predict the insulating nature of SFO. It is evident that significant contributions in the density of states are coming from localized the Sm-$4f$ and the Fe-$3d$; and also from the O-$2p$ orbitals and mixing occurs between these orbitals at $E_F$, see Fig.~\ref{Figure_02}(a), (b) and (c). This embodies the fact that GGA-PBE exchange-correlation functional incorrectly models of the electron interactions between these states.

Now we present more accurate HSE06 hybrid functional based calculations to explain the detail electronic structure of the SFO. At high binding energies of -20 eV, most of the contribution to the DOS comes from the Sm-$5p$ states, see Fig.~\ref{Figure_03}(a). The Fe-$3d$ and Fe-$3p$ states are also present at this energy range and provide small contribution to DOS. From -10 to -2 eV, significant amount of mixing occurs between Sm-$4f$, Fe-$3d$, and O-$2p$ resulting strong hybridization among these states, see Fig.~\ref{Figure_03}(a), (b) and (c). Near the top-of the valence band (VB) within the energy window of -2 to 0 eV ($E_F$), energy bands are almost exclusively derived from hybridization between Fe-$3d$ and O-$2p$ states. The bottom of the conduction band are formed due to mixing between Sm-$4f$, Fe-$3d$ and O-$2p$ states around 2 eV. As we go higher in energy, around 3 eV, the mixing occurs among dominant Fe-$3d$ and small Sm-$4f$ and O-$2p$ states. From 4 to 10 eV energy range, the DOS originates from Sm-$4f$, Sm-$5d$, Fe-$3d$ and O-$2p$ states. Beyond 10 eV, Sm-$5p$, Sm-$4f$, Sm-$5d$, Fe-$3d$, Fe-$3p$ and O-$2p$ mix together to form the DOS.

We perform electronic band structure calculations along the high symmetry directions $\Gamma$, X, R, S, T, U, Y, and Z  in the Brillouin zone of the orthorhombic SFO within the energy range from -20 to 20 eV centered at $E_F$, see Fig.~\ref{Figure_04}. The valence band maximum and the conduction band minimum occur at R point indicating an direct bandgap of 1.75 eV for SFO. It is interesting to note that energy levels near the bottom of the CB has small dispersion which can be attributed to the fact that localized Sm-$4f$ and Fe-$3d$ orbitals mostly responsible for constructing those bands. The dispersion near the top of the valence band is more pronounced which resembles the presence of spatially delocalized O-$2p$ states.

\section{\label{sec:level2} Conclusion}
We have studied the structural, elastic and electronic structure of orthorhombic SFO using the PAW method within the framework of DFT. We performed first principles  calculations for the total energies for different possible SFO magnetic configurations and found G-AFM to be the ground state with minimum energy which is consistent with experimental observations. We have calculated the lattice parameters, atomic positions, relevant ionic radii, bond lengths and bond angles for SFO for a number of standard exchange-correlation functionals and made a comparative study between them. We simulated the elastic properties of the SFO in terms standard parameters like elastic constants and moduli and studied the mechanical stability of the SFO. For electronic structure analysis, simulations have been performed using approximations in the Jacob's ladder for the exact exchange-correlation functional  and a detail discussions are presented for comparative analysis.  

The presence of heavy atom like Sm (atomic number Z = 62) in SFO indicates that the effect of spin–orbit coupling (SOC) can be important to understand the electronic properties in more finer details. This requires inclusion of relativistic treatment through the Dirac Hamiltonian. The core electrons which are moving fast near the heavy nucleus can be treated with full relativistic Dirac equation and the valence electrons are modelled within the scalar approximations. In future the inclusion of relativistic corrections in the DFT calculations to explore the intricate interplay between SOC and electronic properties can provide more detail insights into physical properties of SFO. 

\section*{Acknowledgments}
We gratefully acknowledge Dr. Tapas Debnath, Department of Theoretical and Computational Chemistry, University of Dhaka for providing the license for the Vienna $Ab$ $Initio$ Simulation Package (VASP). We are also very thankful to Dr. Shafiul Alam, Department of Electrical and Electronic Engineering, University of Dhaka for the access to high-performance computing facility.

\section*{Author contributions} 
S.A. and S.S.N. contributed equally. I.A. and A.K. planned the project. 
S.A., S.S.N., A.K.M.S.H.F., T.H. and S.C. performed the simulations. S.A., S.S.N. and I.A. analyzed the data. All authors contributed in the writing of the manuscript.

\textbf{Competing interests}: The authors declare no competing interests.



\bibliographystyle{apsrev4-1}
\bibliography{main}
\end{document}